\documentclass[]{aa}
\usepackage{natbib,psfig,graphicx}
\def\mathnew{\mathsurround=0pt}
\def\simov#1#2{\lower .5pt\vbox{\baselineskip0pt \lineskip-.5pt
\ialign{$\mathnew#1\hfil##\hfil$\crcr#2\crcr\sim\crcr}}}

\def\MeV{Me\kern-0.11em V}
\def\keV{ke\kern-0.11em V}

\begin{document}

\title{The isolated fossil group RXJ1119.7+2126
\thanks{Based on observations made at Observatoire de Haute Provence (CNRS), 
France.
Also based on the use of the NASA/IPAC Extragalactic Database (NED) which is 
operated by the Jet Propulsion Laboratory, California Institute of 
Technology, under contract with the National Aeronautics and Space 
Administration.
}}

\offprints{C. Adami \email{christophe.adami@oamp.fr}}

\author{ C. Adami\inst{1} \and 
D. Russeil\inst{1} \and 
F. Durret\inst{2,3} 
}

\institute{
LAM, Traverse du Siphon, 13012 Marseille, 2 place Le Verrier, 13248 Marseille, 
France
\and
Institut d'Astrophysique de Paris, CNRS, UMR~7095, Universit\'e Pierre et 
Marie Curie, 98bis Bd Arago, 75014 Paris, France 
\and
Observatoire de Paris, LERMA, 61 Av. de l'Observatoire, 75014 Paris, France
}

\date{Accepted . Received ; Draft printed: \today}

\authorrunning{Adami et al.}

\titlerunning{A clue to fossil group evolution}

\abstract
{Fossil groups are galaxy structures that probably underwent a nearly
complete fusion of all intermediate magnitude galaxies into a single
large central dominant galaxy. However, the formation and evolution
processes of these structures are still not well understood. }
{In order to test this scenario and its implications we studied the 
fossil group
RXJ1119.7+2126, based on available spectroscopy of the galaxies in 
the low density large scale region around the fossil group and deep B and 
R band imaging of its close vicinity and three comparison fields.}
{We used spectroscopic data to investigate the degree of isolation
of RXJ1119.7+2126 in terms of bright neighbour galaxies. The imaging
data were used to derive the color magnitude relation and select faint
galaxies statistically belonging to this structure.}
{The structure appears as a very isolated group exhibiting a red sequence in
the color magnitude relation with characteristics close to the red sequences
already observed for other fossil groups.}
{All these results can be interpreted consistently in the framework of
the building up process generally proposed for fossil groups.}  

\keywords{galaxies: groups: individual (RXJ1119.7+2126)} 

\maketitle

\section{Introduction}

Hierarchical building of structures is a key ingredient for
cosmological models, since galaxy structures such as groups and clusters
are expected to be located at the intersections of cosmic filaments
(e.g. Lanzoni et al. 2005 and references therein). These structures
are then continuously fed by infalling field matter (galaxies and
gas). However, a peculiar class of structures seems not to follow this
general behaviour: fossil groups (e.g. Ulmer et al. 2005, Mendes de
Oliveira et al. 2006 and references therein).

These structures are considered as the ultimate stage of group
evolution: the nearly complete fusion of all the bright and
intermediate magnitude galaxies of the group into a single bright
galaxy. This resulting galaxy is brighter than the second remaining
group galaxy (within half the projected virial radius) by at least 2
magnitudes (in the R band). However, the extended X-ray gas envelope 
is still present
and more luminous than 10$^{42}~$h$_{50}^{-2}$ erg~s$^{-1}$ (Jones et
al. 2003). Recent numerical simulations by D'Onghia et al. (2005) have
indeed shown that fossil groups have already assembled 50\% of their
final dark matter mass at $z \geq 1$ and that they subsequently grow
by minor merging. D'Onghia et al. (2005) have also shown that they are
expected to be overluminous in X-rays relatively to non-fossil
groups. It was also suggested in this paper that fossil groups exist
only because infall of $L \sim L_*$ galaxies happens along filaments
with small impact parameters.

We can also explain this lack of bright galaxies if
fossil groups are isolated from any surrounding large scale cosmic 
structure during a time typically longer than the galaxy crossing time
in the considered group (e.g. Sarazin 1986). In order to test this
hypothesis, we investigated in detail the fossil group RXJ1119.7+2126
(RXJ1119 hereafter).

RXJ1119 is one of the less massive known fossil groups (Jones et al.
2003).  It is located at $\alpha$=11$^h$19$^{mn}$43.7$^s$,
$\delta$=21$^\circ$26'50'' (J2000.0) and at a redshift of 0.061. This
low value of the redshift allows its study with moderate size
telescopes and its coordinates are outside any major recent survey
such as the SDSS or 2dF. It is located on the same line of sight
but significantly beyond the well known filaments embedding the Coma
cluster.  Its field of view is therefore heavily polluted by
foreground galaxies.  

In section 2, we present our data and the characteristics deduced for
this structure in terms of bright neighbouring galaxies.  Section 3
concerns the study of the faint galaxy population of RXJ1119.  Section
4 is the conclusion. In this paper we assume H$_0$ = 75 km s$^{-1}$
Mpc$^{-1}$, $\Omega _m$=0.3 and $\Omega _{\Lambda}$=0.7. All
magnitudes are given in the Vega system.

\section{Bright neighbouring galaxies?}

The first goal of this paper is to investigate the isolation degree of
RXJ1119 and the 
surrounding Large Scale Structure (LSS hereafter). The isolation
degree can be understood in terms of bright galaxies. In the model of
a nearly complete fusion of the intermediate magnitude galaxies 
into a single bright galaxy, this galaxy is expected to be the
dominant one of the neighbouring space. The minimal size of the space
to be considered is imposed by the fact that RXJ1119 has to be isolated
and therefore probably located in a void. The average size of the
known voids is close to 40~Mpc in diameter (Hoyle $\&$ Vogeley 2004). 
We therefore tried to detect if there
are brighter galaxies than the central RXJ1119 galaxy in similar
areas around the fossil group. In other words, is the RXJ1119 central
galaxy the brightest of its bubble (if located in such a bubble)?

In order to estimate (in a 7$\times$7~deg$^2$ area or 
$\sim$28$\times$28 Mpc$^2$, see below) the expected number 
of galaxies at z$\sim$0.061 brighter 
than the RXJ1119 central galaxy (${\rm M_R} = -22$ in our cosmology: 
Jones et al. 2003) and twice as bright as the RXJ1119 central galaxy 
(${\rm M_R} = -22.75$), we used the field luminosity functions of Ilbert et
al. (2005) and the cluster luminosity functions of Popesso et
al. (2005).  On the one hand, without any rich galaxy structures in the
bubble, the field luminosity functions predict respectively about
50 galaxies per magnitude bin brighter than ${\rm M_R} = -22$ in our
sampled volume and less than 5 galaxies per magnitude bin brighter
than ${\rm M_R} = -22.75$. On the other hand, if one or several rich 
clusters are present in the area, these numbers increase significantly: 
several dozen galaxies brighter than ${\rm M_R} = -22.75$ would be expected 
per magnitude bin.

\begin{figure}
\centering
\mbox{\psfig{figure=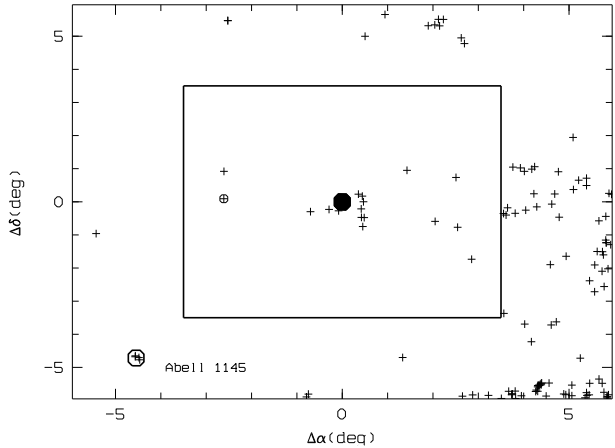,width=9cm,angle=270}}
\caption[]{Map of the 12$\times$12~deg$^2$ area around RXJ1119 given
in positions relative to the center of RXJ1119. The large rectangle is
the inner 7$\times$7~deg$^2$ area. Crosses are the galaxies with a
redshift in the [0.051,0.071] redshift range. Inside the
7$\times$7~deg$^2$ area, the only galaxy brighter than the RXJ1119
central galaxy by more than 0.2 magnitude is shown as the small
circled cross. The position of Abell~1145 is given at the bottom left
of the plot. RXJ1119 is shown by the central filled circle.}
\label{fig:fig1}
\end{figure}

In order to compare observations with these predictions, we compiled (from NED) all
the redshifts known in a 12$\times$12~deg$^2$ area ($\sim$47$\times$47
Mpc$^2$) whatever their magnitude.  We limited the searched redshift
range to [0.051,0.071]. This range corresponds to a physical size of
$\sim$90 Mpc and $\pm$3 times the galaxy velocity dispersion of a
massive cluster. In this range, there are 20 galaxies in a
7$\times$7~deg$^2$ area (the area in which RXJ1119 seems isolated,
i.e. $\sim$28$\times$28 Mpc$^2$) and 143 in the 12$\times$12~deg$^2$
area. The closest known cluster is Abell~1145, only included in the
outer region. We show these positions in Fig.~\ref{fig:fig1} along
with the two searched areas. Now, restricting this list to galaxies
brighter than the central RXJ1119 galaxy (magnitudes taken from the
NED database), in the 7$\times$7~deg$^2$ area, there is only one
galaxy (as compared to 50 from the expected numbers even without any
rich galaxy structure in the field). This galaxy (also brighter than
${\rm M_R} = -22.75$) is, moreover, quite distant from the fossil
group (about 12~Mpc or slightly more than 3 deg). The redshift catalog
is however not complete down to ${\rm M_R} = -22$ and we cannot
exclude that a significant number of galaxies with unknown redshift
are in the considered redshift range.

We then measured additional redshifts for 20 of the 23 galaxies
extracted from NED (without the redshift information) in the largest
area down to ${\rm M_R} = -22.75$ (plus one fainter galaxy that we
caught at the same time as a brighter target). This produced a
spectroscopic catalog $\sim$100$\%$ complete down to this magnitude.
These spectroscopic observations were made at the Observatoire de
Haute Provence with the 193~cm telescope and the Carelec
spectrograph. We used the 133 \AA /mm configuration (resolution R=900)
and the useful spectral range was 3700--6700~\AA. This instrument is a
single slit device well adapted to such a survey: the target density
is 0.16 per deg$^2$ making high multiplex multi-object spectrographs
inefficient. Results are given in Table~\ref{tab:tab1}.

None of these additional galaxies is in the [0.051,0.071] redshift
range: most are located around z=0.023, and 4 are around
z=0.035. Therefore there is only one galaxy with ${\rm M_R} \leq
-22.75$ galaxies at the RXJ1119 redshift and in the 7$\times$7~deg$^2$
area (assuming that the 3 galaxies with unknown redshift are not at
z=0.061). This is clearly lower than the several dozen of galaxies
expected if rich galaxy structures were present. This is even lower
than the expected galaxy number assuming a ``normal'' field galaxy
population.

RXJ1119 therefore seems to be very isolated in terms of bright galaxies 
and it appears to be the dominant structure of its
bubble. This bubble would have a typical diameter of 28 Mpc, close to
the known void dimensions (Hoyle $\&$ Vogeley 2004). However, this
remains to be confirmed for faint galaxies with a redshift catalog complete 
down to fainter magnitudes than the one we have. For example, only less than 10$\%$ of all
galaxies known in the 7$\times$7~deg$^2$ area (from NED) have a measured 
redshift.

\begin{table}
\caption{Col.~1: NED galaxy name; col.~2: measured redshift; col.~3: nature
of the spectrum: emission lines (Em), absorption lines (Abs). The typical
redshift uncertainty for these galaxies is 0.001.}
\begin{tabular}{lll}
\hline
Name & z & Spectrum \\ 
\hline
CGCG 125-023  & 0.023 & Em + Abs: Em dominated \\
UGC 06172  & 0.021 & Em + Abs: Em dominated \\
CGCG 096-008  & 0.026 & Abs only \\
CGCG 096-009  & 0.027 & Abs only \\
MCG +03-29-009  & 0.027 & Abs only \\
KUG 1110+235A  & 0.023 & Em + Abs: Em dominated \\
KUG 1110+235  & 0.026 & Em + Abs: Em dominated \\
KUG 1112+236A   & 0.022 & Em + Abs: Em dominated \\
CGCG 096-016  & 0.023 & Em + Abs: Abs dominated \\
CGCG 126-014  & 0.023 & Em + Abs: Em dominated \\
CGCG 126-018  & 0.023 & Abs only \\
UGC 06301  & 0.035 & Em + Abs: Em dominated \\
KUG 1115+236  & 0.022 & Abs only \\
CGCG 126-026  & 0.027 & Em + Abs \\
UGC 06336  & 0.026 & Abs only \\
CGCG 126-047  & 0.023 & Em + Abs: Abs dominated \\
CGCG 126-052  & 0.041 & Em + Abs: Em dominated \\
CGCG 126-055  & 0.024 & Em + Abs: Abs dominated \\
CGCG 126-076  & 0.032 & Em + Abs: Em dominated \\
KUG 1130+249A & 0.024 & Em + Abs: Em dominated \\
CGCG 126-088  & 0.034 & Abs only \\
\hline
\end{tabular}
\label{tab:tab1}
\end{table}

\section{The RXJ1119 faint galaxy population}

\subsection{Imaging data}

\begin{figure}
\centering
\mbox{\psfig{figure=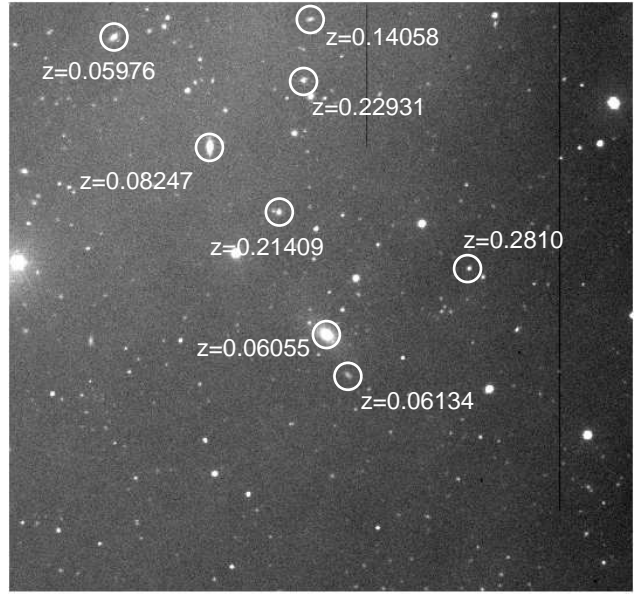,height=8.cm,angle=0}}
\caption[]{R band 10.2$\times$9.7~arcmin$^2$ image of RXJ1119 with the
known redshifts in the field.}
\label{fig:fig2first}
\end{figure}

\begin{figure}
\centering
\mbox{\psfig{figure=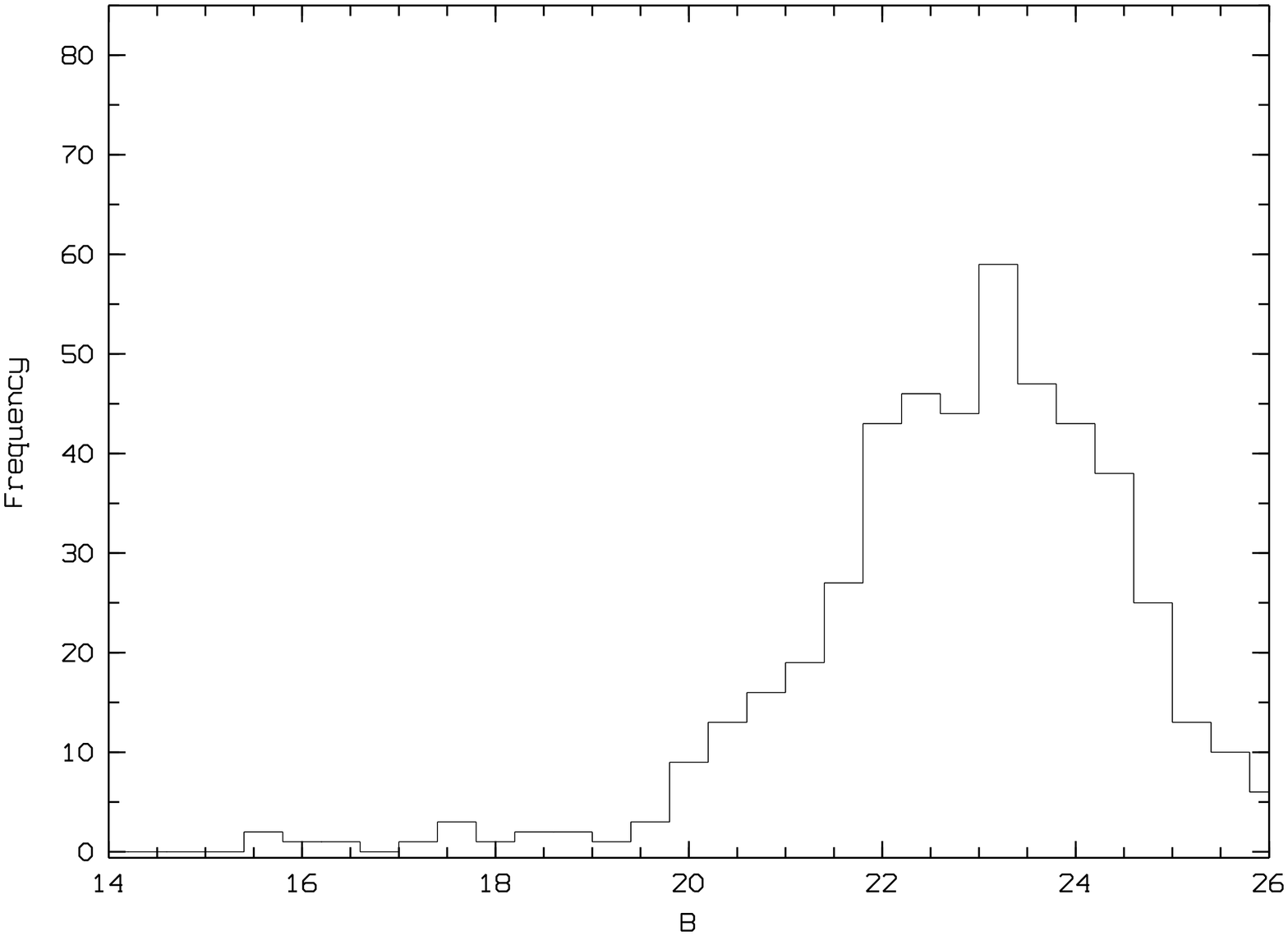,height=6.cm,angle=270}}
\mbox{\psfig{figure=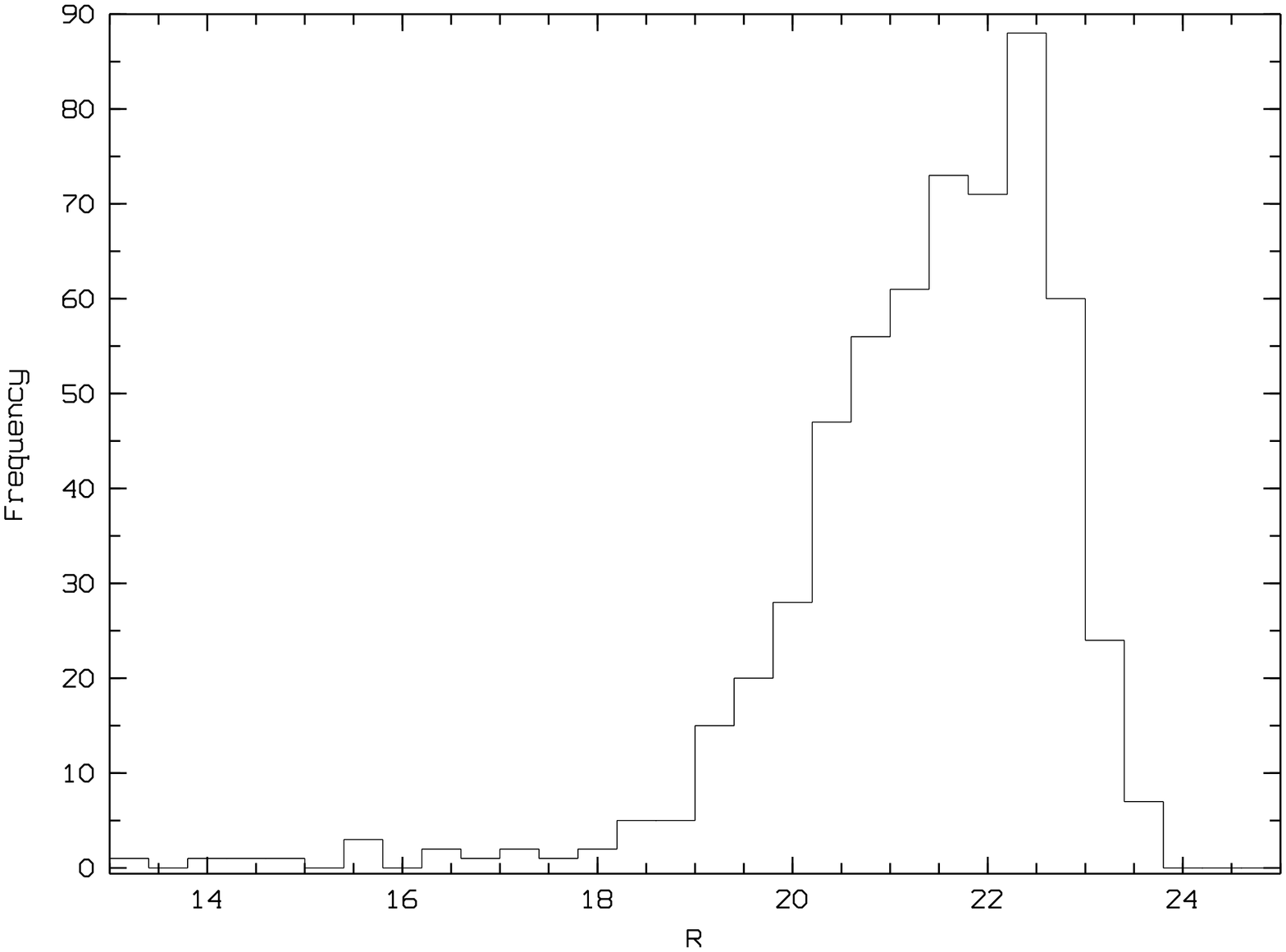,height=6.cm,angle=270}}
\caption[]{Magnitude histograms of galaxies along the RXJ1119 
line of sight. Top: B band, bottom: R band.}
\label{fig:fig2}
\end{figure}

We investigate in this section what happens at smaller scales,
as the second goal of this work is to study the faint galaxy
population in RXJ1119 itself. Only little is known on the galaxy
populations in fossil groups beyond the spectroscopic limit and the
second brightest galaxy for most of these structures (see however
Mendes de Oliveira et al. 2006). Rather than measuring redshifts of
galaxies of this population, we chose to obtain deep B and R band
imaging data in order to draw a color magnitude relation (CMR
hereafter). This was done at the OHP 120~cm telescope. A CCD camera
with a $\sim 11.5 \times 11.5$~arcmin$^2$ field (1024$\times$1024
pixels with a pixel size of 0.69$\times$0.69 arcsec$^2$) is mounted on
this telescope. We used Johnson B and R filters and we observed RXJ1119
itself as well as three comparison fields (see
Table~\ref{tab:tab0}). The comparison fields (C1, C2 and C3) were
chosen in order to have a Galactic extinction similar to that of RXJ1119 (from 
the Schlegel et al. 1998 maps) and not to include nearby known galaxy
structures. Data were acquired under photometric conditions and
seeing of the order of 2.5 arcsec. The final useful area for
RXJ1119 with both B and R data available was $\sim 10.2 \times
9.7$~arcmin$^2$. The three comparison fields cover 336 arcmin$^2$
in total and are distributed in a region of $\sim$45$\times$45 deg$^2$ on 
the sky.  The images
were calibrated using Landolt (1992) standard stars. We extracted
catalogs of objects from these images using the SExtractor package
(Bertin $\&$ Arnouts 1996). An R image of the RXJ1119 field is
displayed in Fig.~\ref{fig:fig2first} and the magnitude histograms are
given in Fig.~\ref{fig:fig2}.

\begin{table*}
\caption{Col.~1: observed field ; col.~2: coordinates (2000); col.~3: 
R exposure time; col.~4: B exposure time; col.~5: E(B-V) following 
Schlegel et al. (1998).}
\begin{tabular}{lllll}
\hline
field & coordinates (2000) & R & B & E(B-V) \\ 
\hline
RXJ1119  & 11 19 43.7, 21 26 50 & 4500s & 10800s & 0.021$\pm$0.001 \\
C1  & 23 40 41.6, 15 07 11 & 4500s & 10800s & 0.026$\pm$0.001 \\
C2  & 01 08 02.1, 12 56 27 & 3600s & 9720s & 0.019$\pm$0.002 \\
C3  & 02 25 49.3, 02 09 45 & 3240s & 8400s & 0.026$\pm$0.001 \\
\hline
\end{tabular}
\label{tab:tab0}
\end{table*}

\begin{figure}
\centering
\mbox{\psfig{figure=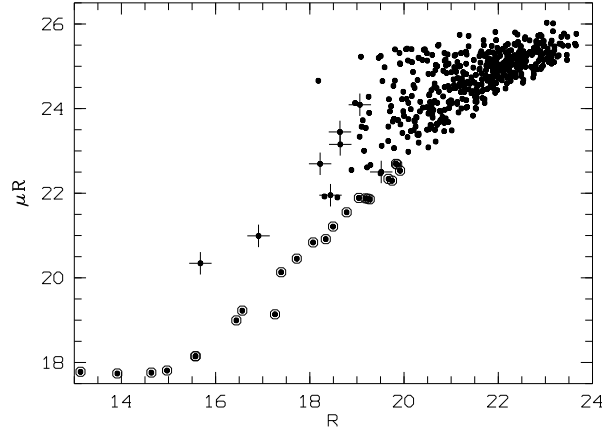,height=6cm,angle=270}}
\caption[]{ R central surface brightness versus total R
magnitude for the RXJ1119 field of view. Objects classified as 
stars are circled dots. We can see
the saturation of the brightest stars. Crossed dots are objects
classified as galaxies using spectroscopy. }
\label{fig:fig3}
\end{figure}

It is now important to perform a star-galaxy separation in order to
avoid star pollution and to take into account the fact that the four
  observed fields do not have the same Galactic latitude. 
Seeing conditions are not good enough to use the
classical SExtractor star-galaxy separation flag. We used instead a
total magnitude vs. central surface brightness plot (e.g. Adami et
al. 2006a) based on the deepest band data (the R band). Results are
  illustrated in
Fig.~\ref{fig:fig3} for RXJ1119, where we clearly see the star 
sequence. We were
therefore able to make an efficient star-galaxy separation down to
R=20 for the four observed fields. For fainter 
magnitudes, all objects are seeing dominated and any
star-galaxy separation would be very uncertain. However the number of
stars fainter than R=20 is small (Gazelle et al. 1995) and we decided
to assume that all objects fainter than this limit are galaxies.

In order to test the separation down to R=20 for RXJ1119, we 
retrieved in NED the 8
galaxies in the imaging field with a known redshift. We see that they
are all classified as galaxies by our method in Fig.~\ref{fig:fig3},
including a relatively compact object.

\subsection{The RXJ1119 Color Magnitude Relation}

We compute here the galaxy distribution in the R/B-R plane for
the RXJ1119 structure. In order to achieve this goal, we have to
subtract the fore and background contributions along the line of
sight to RXJ1119. We used a technique similar to that described in
Adami et al. (2006b). In a few words, the fore and background RXJ1119
line of sight galaxy populations were estimated using the three
comparison fields.  For a given R and B-R, the number of galaxies per
deg$^2$ in the RXJ1119 structure (N$_{RXJ1119}$) was computed with the
number of galaxies per deg$^2$ along the RXJ1119 line of sight
(N$_{RXJ1119\ l.o.s}$) and the numbers of galaxies per deg$^2$ along the
C1, C2 and C2 lines of sight (N$_{C123\ l.o.s}$) as:

N$_{RXJ1119}$(R,B-R) = N$_{RXJ1119\ l.o.s}$ - N$_{C123\ l.o.s}$

This assumed that the C1, C2 and C3 fields were representative of the global
galaxy population. In order to estimate the uncertainty on the previous
difference, we computed the standard deviation
$\sigma$(R,B-R) of the counts (for given R and B-R)
between the three fields. This gave us an R/B-R map of the uncertainty.

Finally, we divided the N$_{RXJ1119}$(R,B-R) map by the $\sigma$(R,B-R) map to
generate a significance map.

\begin{figure}
\centering
\mbox{\psfig{figure=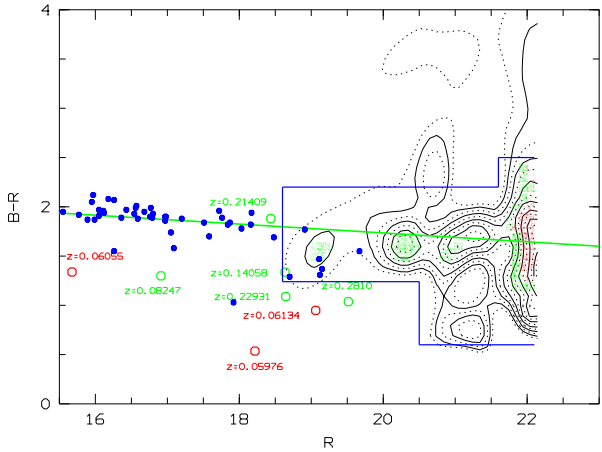,width=9.cm,angle=270}}
\caption[]{Total R magnitude versus B$-$R color for galaxies
statistically inside the RXJ1119 structure. Galaxies with a known
redshift are circles (red: inside the structure, green:
outside). Using an adaptative kernel technique, we overplotted the
density contours in the R/B$-$R space. The green line has been
normalized to fit the RXJ1552 and RXJ1416 galaxies in the B-R/R space
and to have the Coma cluster slope.  Blue dots are galaxies with
spectroscopy from the RXJ1552 and RXJ1416 fossil groups (Mendes de
Oliveira et al. 2006, Cypriano et al. 2006) with their magnitudes and
colors translated in our system. The blue-delimited area corresponds
to our modelling of the positive loci of the density contours. We
limit the graph at R$\sim$22 to take into account the completeness
limit of our catalogs.}
\label{fig:fig4}
\end{figure}

The resulting CMR is shown in Fig.~\ref{fig:fig4}. We only plotted
  contours of the areas where N$_{RXJ1119\ l.o.s}$ was greater than 
N$_{C123\ l.o.s}$, that is the places where galaxies are statistically present 
inside the RXJ1119 structure.

First, this figure shows that RXJ1119 is indeed a fossil group
(besides the fact that its X-ray luminosity estimated by Jones et
al. (2003) is in the allowed range for such groups): the magnitude
difference between the brightest and second brightest galaxies in the
structure is larger than 2 magnitudes. The magnitude difference
between the brightest galaxy and the bright galaxy peak (at R$\sim$19
and B-R$\sim$1.6) in Fig.~\ref{fig:fig3} is also larger than 2
magnitudes. The only galaxy with a magnitude difference smaller than 2
is at a redshift of $\sim$0.0825, and does not belong to the
structure.

Second, the presence of a Red Sequence (RS hereafter) in the CMR is a
well known characteristic of massive galaxy systems (see e.g. Godwin $\&$
Peach 1977 or Mazure et al. 1988). The usual negative slope of this RS is a
metallicity effect (Kodama $\&$ Arimoto 1997) originating from the
higher ability of massive early type galaxies (as opposed to low mass
objects) to keep metals against dissipative processes as supernovae,
and then to form redder stars. However, only little is known on a
possible RS in fossil groups and especially in RXJ1119.  
Fig.~\ref{fig:fig4} shows concentrations of galaxies that are similar 
to a RS. The green shaded areas are places where the RXJ1119 structure counts
are significant between the 2$\sigma$ and 3$\sigma$ levels. The red shaded
areas are significant above the 3$\sigma$ level.

In order to compare with other fossil groups, we superposed on
this figure the CMR of RXJ1552.2+2013 (z=0.136, RXJ1552 hereafter) and
RXJ1416.4+2315 (z=0.137, RXJ1416 hereafter, Mendes de Oliveira et
al. 2006 and Cypriano et al. 2006). We used Fukugita et al. (1995) to
take into account the redshift and filter differences and
translated their AB magnitudes into the Vega system. The green line
shown in Fig.~\ref{fig:fig4} has been chosen to fit the RXJ1552 and
RXJ1416 galaxies in the B-R/R space and to have a slope of $-0.045$ (see
Adami et al. 2006b). The RXJ1119 overall RS is in good agreement with
those of RXJ1552 and RXJ1416. A remarkable characteristic of
RXJ1119 compared to RXJ1552 and RXJ1416 is that the population of
galaxies is very poor. Translated to z=0.061, the RXJ1416 and RXJ1552
fields show a significant population of galaxies between R=16 and R=19
while the first significant RXJ1119 galaxy concentrations occur for R
fainter than 18.5. This is probably related to the fact that RXJ1416
and RXJ1552 are much more massive than RXJ1119 (from the X-ray
luminosities given by Jones et al. 2003) and probably underwent a
larger field galaxy infall, which compensated more efficiently the
galaxy structure depopulation by merging or disruptions. We also note
that the brightest galaxy of RXJ1119 is bluer than the mean RS, as
already observed in RXJ1552 and RXJ1416 (Mendes de Oliveira et
al. 2006, Cypriano et al. 2006).

\begin{figure}
\centering
\mbox{\psfig{figure=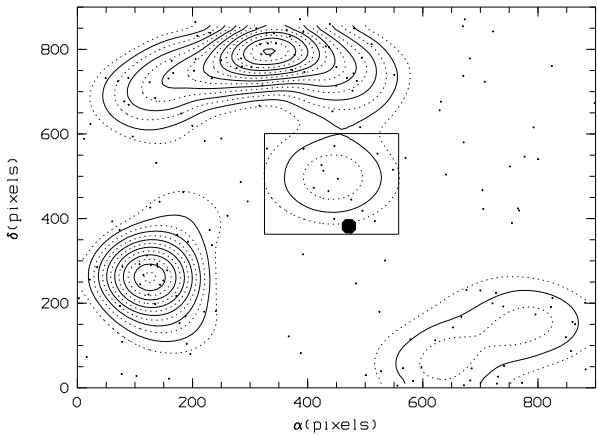,height=6.cm,angle=270}}
\caption[]{Adaptative kernel galaxy density map (in pixels) for
galaxies inside the blue-delimited area in Fig.~\ref{fig:fig4}. 
The X-ray position of RXJ1119 is shown by the
filled circle. }
\label{fig:fig5}
\end{figure}

We can now use the RS in the RXJ1119 CMR to select galaxies
possibly in RXJ1119 and to compute the galaxy density map of the
structure. This is done by selecting galaxies inside the
blue-delimited region (arbitrarily defined to encompass the positive
areas) in Fig.~\ref{fig:fig4}. This does not totally prevent us from
selecting background galaxies with the same color/magnitude
combination, but the contrast between structure and field galaxies is
increased by a factor of about 3.  We computed such a galaxy density
map in Fig.~\ref{fig:fig5}. This figure shows that there is a
galaxy concentration close to the X-ray position of RXJ1119 (but not
centered on it) and with an extension larger than the X-ray halo (see
Jones et al. 2003). In order to favor a possible spectroscopic survey
of these faint galaxies, we give their list in Table~\ref{tab:tab3}.

\begin{table}
\caption{Coordinates, R band magnitude and B-R colors of objects included in
  the central concentration around RXJ1119. They were previously
  uncataloged in NED and there is no redshift available for these objects.}
\begin{tabular}{lll}
\hline
Coordinates & R & B-R \\ 
\hline
11:19:50.8, +21:29:03.0 & 21.84 & 1.33 \\
11:19:41.1, +21:28:56.0 & 21.66 & 1.33 \\
11:19:46.0, +21:28:36.6 & 21.67 & 1.92 \\
11:19:41.2, +21:27:03.9 & 21.02 & 1.10 \\
11:19:45.3, +21:27:53.2 & 20.91 & 1.96 \\
11:19:46.6, +21:27:58.6 & 21.20 & 1.73 \\
11:19:44.9, +21:29:08.0 & 21.21 & 1.28 \\
11:19:45.8, +21:28:27.0 & 21.14 & 1.90 \\
11:19:44.8, +21:27:07.2 & 19.08 & 1.67 \\
11:19:42.3, +21:27:21.0 & 19.25 & 1.41 \\
11:19:47.6, +21:29:03.2 & 19.96 & 1.78 \\
11:19:43.3, +21:27:39.3 & 20.21 & 2.19 \\
11:19:39.7, +21:28:29.3 & 20.03 & 1.67 \\
11:19:44.5, +21:28:13.8 & 20.11 & 1.91 \\
\hline
\end{tabular}
\label{tab:tab3}
\end{table}

This concentration is isolated from several other concentrations of galaxies
sharing the same colors. Assuming that these galaxies are part of the RXJ1119
structure, it therefore seems that the inner core of RXJ1119 
is also isolated in terms of faint galaxies. The closer external galaxy
  concentrations are located at a mean distance of $\sim$150 kpc from 
the central structure.

Using a bidimensional Kolmogorov-Smirnov test, we show that the
galaxies inside and outside the central galaxy concentration of
Fig.~\ref{fig:fig5} (and inside the blue-delimited area of
Fig.~\ref{fig:fig4}) do not have a different distribution inside the
R/B-R space (at the 99$\%$ level). Therefore they probably followed
similar evolutionary paths. The galaxies very close to the RXJ1119
center and at the edges of the structure can therefore be considered
as similar.

\section{Conclusions}

Let us summarize our main findings:

- RXJ1119 is located at the center of a low galaxy density bubble. 

- RXJ1119 is a fossil group which shows a RS in its CMR, quite
similar to RSs of other fossil groups.

- RXJ1119 is a very poor and low mass structure (even compared to other 
fossil groups) with the first galaxies statistically inside the structure 
fainter than R=18.5.

 - The faint RXJ1119 galaxy population is strongly clustered and
isolated from surrounding galaxy layers. RXJ1119 is therefore isolated
in terms of bright and faint galaxies, even at small scales. This
{\sl empty ring} seen in Fig.~\ref{fig:fig5} has a radius of 150 to 200
kpc. This could be interpreted as the radius where galaxies start to
be significantly influenced by the potential well of the fossil
group. However, using the formulae relating the virial radius to the
X-ray temperature T of the structure (Jones et al. 2003) gives a very
low estimate of T (only a fraction of a keV). An estimate of T based
on X-ray data is therefore needed to confirm or invalidate this
interpretation.

Overall, these conclusions are in good agreement with the following building
scenario for fossil groups: an isolated galaxy
structure passively evolving with only minor field contribution due to
its location in a low density environment (see also D'Onghia et al. 2005 and
references therein). We now need to put this result on firmer
grounds by analysing in a similar way the neighbourhood of other known
fossil groups.

\begin{acknowledgements}
The authors thank the referee for very useful and constructive comments.
The authors are grateful to the OHP team and to the students of the 2005/2006
promotion of the Aix-Marseille I Rayonnement, Plasmas et Astrophysique M2. 
\end{acknowledgements}

\end{document}